\documentclass[manuscript]{emulateapj}

\newcommand{\LX}{L_{\rm X}}

\newcommand{\Mw}{\dot{M}_{\rm w}}

\newcommand{\percc}{\rm \,cm^{-3}}

\newcommand{\ps}{{\rm s}^{-1}}

\newcommand{\kmps}{{\rm km}\,{\rm s}^{-1}}

\newcommand{\ergps}{{\rm erg}\,{\rm s}^{-1}}
\newcommand{\bcen}{\begin{center}}
\newcommand{\ecen}{\end{center}}
\newcommand{\be}{\begin{equation}}
\newcommand{\ee}{\end{equation}}
\newcommand{\bdis}{\begin{displaymath}}
\newcommand{\edis}{\end{displaymath}}
\newcommand{\NeII}{Ne\,{\sc ii}}
\newcommand{\NeIII}{Ne\,{\sc iii}}
\newcommand{\OI}{O\,{\sc i}}

\slugcomment{Received 2009 November 18; accepted 2010 March 25; published 2010 April 22}

\begin{document}

\title{\NeII\ Fine-Structure Line Emission from the Outflows of Young Stellar Objects}

\author{Hsien Shang\altaffilmark{1}, Alfred E. Glassgold\altaffilmark{2},
Wei-Chieh Lin\altaffilmark{1,3}, and Chun-Fan J. Liu\altaffilmark{1,3}}

\altaffiltext{1}{Institute of Astronomy and Astrophysics (ASIAA),
and Theoretical Institute for Advanced Research in Astrophysics
(TIARA), Academia Sinica, P. O. Box 23-131, Taipei 10641, Taiwan}
\altaffiltext{2}{Astronomy Department, University of California,
Berkeley, CA 94720-3411, USA} \altaffiltext{3}{Graduate Institute of
Astronomy and Astrophysics, National Taiwan University, No.~1,
Sec.~4, Roosevelt Road, Taipei 10617, Taiwan}

\begin{abstract}

The flux and line shape of the fine-structure transitions of \NeII\
and \NeIII\ at 12.8 and 15.55\,$\mu$m and of the forbidden transitions
of \OI\ $\lambda6300$ are calculated for young stellar objects with a
range of mass-loss rates and X-ray luminosities using the X-wind model
of jets and the associated wide-angle winds. For moderate and high accretion
rates, the calculated \NeII\ line luminosity is comparable to or much
larger than produced in X-ray irradiated disk models. All of the line
luminosities correlate well with the main parameter in the X-wind model,
the mass-loss rate, and also with the assumed X-ray luminosity --- and with
one another. The line shapes of an approaching jet are broad and have strong 
blue-shifted peaks near the effective terminal velocity of the jet. They 
serve as a characteristic and testable aspect of jet production of the neon 
fine-structure lines and the \OI\ forbidden transitions.

\end{abstract}


\section{Introduction}

More than 50 detections of the \NeII\ 12.8\,$\mu$m line have been
made in young stellar objects (YSOs) by the {\it Spitzer Space Telescope} 
(Espaillat et al.~2007; Lahuis et al.~2007; Pascucci et al.~2007; Ratzka et al.~2007;
Carr \& Najita~2008) and by ground based telescopes (Herczeg et al.~2007;
Najita et al.~2009; van Boekel et al.~2009; Pascucci \& Sterzik 2009).
The neon fine-structure lines were predicted by Glassgold et al.
 (2007; henceforth GNI07) to arise in disk atmospheres
irradiated by X-rays which ionize neon and generate the warm
and high-ionization conditions needed to excite the lines. However,
the formation of even a single low-mass star occurs in a multi-component
system consisting of a collapsing cloud core, a disk, an accretion
funnel and outflows arising close to the star. Thus various authors have
suggested that the \NeII\ 12.8\,$\mu$m line might be generated elsewhere
in the system and not just by the disk (e.g., Meijerink et al. 2008,
henceforth MGN08; Alexander 2008; van Boekel et al.~2009;
Najita et al.~2009; Flaccomio et al. 2009; Guedel et al. 2010).

Information on the origin of the \NeII\ emission can be obtained from the 
velocity resolved profiles of the 12.8\,$\mu$m line. Najita et al.~(2009) 
discussed four such measurements: TW Hya (Herczeg et al.~2007); T Tau 
(van Boekel et al.~2009); AA Tau, and GM Aur (Najita et al.~2009). TW Hya 
and GM Tau are transitional disks, the former seen almost face-on and the 
latter at an inclination of $54 \arcdeg$, whereas AA Tau is a classical 
T Tauri star (TTS) with a large inclination angle ($75 \arcdeg$). Pascucci 
\& Sterzik (2009) have recently reported four more line shape measurements: 
TW Hya, T Cha, Sz 73, and CS Cha; T Cha, and CS Cha are also transitional 
disks. Najita et al.~(2009) concluded that the line profiles for TW Hya, 
AA Tau, and GM Aur are consistent with the emission arising mainly from a 
gravitationally bound disk atmosphere. Pascucci \& Sterzik (2009) measured 
small blue shifts in three transition disks which they interpreted as arising 
from low-velocity ($\sim 10\, \kmps$) photo-evaporative outflows, 
following Alexander (2008). Observations of T~Tau by van Boekel et al.~(2009)
show that the \NeII\ 12.8\,$\mu$m emission in this case arises from more than 
one location within a complicated and only partially revealed triplet stellar 
system. T Tau is one of several accreting sources with very strong \NeII\ 
emission (Guedel et al.~2008). Its total \NeII\ luminosity is 20 times larger
than measured for many revealed TTSs observed by {\it Spitzer}. The line
shape of Sz 73 shows mainly blue-shifted emission extending well beyond 
$-100\, \kmps$, indicative of a strong stellar outflow. Neufeld et al.~(2006) 
detected the \NeII\ 12.8\,$\mu$m line in HH objects in the HH 7-11 outflow.

We show here that jets from YSOs, long known to emit the forbidden
optical lines of heavy atomic ions, also generate the mid-infrared
transitions of \NeII\ and \NeIII. Using the X-wind model, we calculate
the luminosities and shapes of these lines and also the forbidden \OI\
$\lambda6300$ transitions for comparison. We obtain results for classical
TTSs (Class I and II YSOs) with varying mass-loss rates, and we
show that active jets can dominate the observed flux of the \NeII\ line.
We also show that the line profiles have a distinctive shape that offers
observational opportunities to verify the origin of the emission in strong
outflow sources and to test the predictions of the X-wind model of jets.

\section{Methodology}

The X-wind theory of low-mass star formation (Shu et al.~1994) describes 
how the interaction of the magnetosphere and the accretion disk truncates 
the disk and forces its inner edge to co-rotate with the star. The interaction 
also drives an out-flowing wind and in-flowing accretion streams. In steady 
state, the inner edge of the disk coincides with the gravitational ``X-point'' 
of the system. Assuming that the flows emanate from this location allows a 
semi-analytic theory to be developed that yields a clear picture of outflows 
from low-mass YSOs. Shu et al.~(1995) showed that the density asymptotes to 
cylindrical contours while the streamlines become radial. The physical conditions 
(temperature and ionization) of the X-wind were developed by Shang et al.~
(2002; henceforth SGSL) using Shang's (1998) semi-analytic representation of 
the flow. SGSL assumed the gas is atomic and heated by shocks and ionized by X-rays.  
They were able to give a good account of the optical observations of the forbidden 
lines and of the radio continuum observations of the Class I source, L1551 IRS~5 
(Shang et al.~2004).

\begin{deluxetable}{lccc}
\tabletypesize{\small}
\tablecaption{Parameters of the Reference Model\label{tbl-stellar}}
\tablehead{\colhead{Parameter}&\colhead{Symbol}&\colhead{}&\colhead{Reference Case}}
\startdata
Stellar mass & $M_\ast$ &&  $0.8\,M_{\odot}$    \\
Stellar radius &$R_\ast$    &&  $3.0 \,R_{\odot}$   \\
Disk truncation radius &$R_{\rm x}$     & &      $4.8 \,R_\ast$ \\
Stellar rotation period &$2 \pi / \Omega_{\rm x}$ & &      7.5\,d       \\
Wind mass-loss rate & ${\dot M}_{\rm w}$ & & $1.6 \times 10^{-8}\,M_{\odot}\,{\rm yr}^{-1}$ \\
Disk accretion rate & ${\dot M}_{\rm d}$ & & $1.2 \times 10^{-7}\,M_{\odot}\,{\rm yr}^{-1}$ \\
Mean terminal wind velocity &${\bar v}_{\rm w}$  & & 195 km s$^{-1}$     \\
Stellar luminosity & $L_\ast$   && $2\,L_{\odot}$       \\
X-Ray luminosity  & $\LX$        && $3\times 10^{31}\,{\rm erg}\,{\rm s}^{-1}$ \\
Heating coefficient &$\alpha_h$   && $2\times 10^{-3}$
\enddata
\end{deluxetable}

In addition to X-rays, SGSL considered a variety of heating and
ionization sources (listed in their Table 1). For example, in
addition to X-rays, they treated H$^-$ photo-detachment, Balmer
continuum photoionization by stellar photons and by UV radiation
from accretion-funnel hot spots, as well as collisional ionization.
The X-rays can ionize a large part of the flow because radiative
recombination occurs on a much longer time scale than the flow
time scale (Bacciotti et al.~1995). Thus a significant level of
ionization generated near the source of the flow is frozen into the
wind out to larger distances. Many heating and cooling processes were
considered, but SGSL found that the most important were: (1)
adiabatic cooling of the expanding flow and (2) heating via the
dissipation of internal velocity fluctuations and shocks. SGSL
formulated the latter with a phenomenological formula for the
mechanical heating rate per unit volume,
\be
\label{mech_heat}
\Gamma=\alpha_h \frac{\rho v^3}{s},
\ee
where $\rho$ and $v$ are the local gas density and flow velocity,
$s$ is the distance the fluid element has traveled along a streamline
to the point of interest, and $\alpha_h \ll 1$ is a phenomenological
coefficient that characterizes the magnitude of the mechanical
heating. Equation~(\ref{mech_heat}) may be considered a prescription
for shock heating in the jet. If velocity fluctuations of magnitude
$\delta v$ generate the mechanical heating, then
$\alpha_h v^3\sim (\delta v)^3$, and the values of $\alpha_h$ needed
to heat the flow, $\sim 0.001$, correspond to moderate velocity
fluctuations, $\delta v/v \sim 0.1$. Values of this order may be
inferred from high-spatial resolution observations of the optical
forbidden lines emitted by revealed jets (e.g., Bacciotti et al.~2000;
Woitas et al.~2002).

We use the SGSL model to calculate the neon fine-structure line emission
for the wind and jet of a low-mass YSO.  We also compare the neon
lines with \OI\ $\lambda6300$, representative of optically forbidden lines.
The results depend on the model parameters listed in Table~1. The most 
important are the wind mass-loss rate and X-ray luminosity. In Table~1, 
both the wind mass-loss rate $\Mw$ and the X-ray luminosity $\LX$ are for 
a one-sided jet. The numerical values in the table are based on prior studies 
of jets (SGSL; Shang et al.~2004). We use the fiducial case of SGSL as our own
reference model for a revealed T Tauri source that undergoes active accretion
and drives a bright optical jet.

SGSL conceived the X-ray emission to arise from a soft coronal source
of dimension $R_\ast$ enhanced by magnetic reconnection regions, one
located in the disk midplane and two others associated with magnetic
field $Y$-configurations above and below the plane (see the schematic
drawing of Shu et al.~1997). They idealized this complex as a series of
point sources, one at the origin and the others displaced along the axis
at heights $\pm R_{\rm X}$ with  $R_{\rm X}\sim 0.07 $\,AU. The total X-ray
luminosity of each source was composed equally of soft and hard X-rays,
described respectively by thermal spectra with
$T_{\rm X} = 1$\,keV and $T_{\rm X} = 2$\,keV.
Because of the small dimensions of all of the sources ($\sim R_{\rm X}$),
they behave much like a single coronal X-ray source with an additional 
hard component. The soft X-rays are absorbed over relatively short distances 
for outflows with appreciable mass loss. According to the reference model, the
more penetrating hard X-rays irradiate one hemisphere with a luminosity,
$\LX \sim 1.5\times 10^{31}\, \ergps$.

The relatively large fiducial value of the X-ray luminosity in Table 1 was
adopted by SGSL to treat active YSOs with bright jets and large accretion
rates. Their conception of the X-ray properties of YSOs was formed by the
results from the early X-ray observatories, {\it ASCA} and {\it ROSAT}, 
as discussed by Shu et al.~(1997) who related the properties of the fluctuating
X-wind model and the emission of X-rays. Following Shu et al.~(1997), SGSL
based their X-ray luminosity on {\it ROSAT} observations of embedded sources
(Neuh\"auser 1997), and took into account the likely effect of absorbing
material close to the YSO that reduces the fraction of X-rays that
escape to be detected by an X-ray observatory. Potentially an even
more important absorber than the wind is the system of accretion columns,
which carry a larger mass flux (e.g., Alexander et al. 2004, 2005).
Screening by circumstellar material may help explain the long-standing
puzzle concerning the larger X-ray luminosities of weak-lined versus classical
TTSs, as suggested earlier by Gahm (1980) and by Walter \& Kuhi (1981).
Gregory et al.~(2007) have demonstrated this effect by calculating the
propagation of X-rays through the accretion funnels of a magnetospheric
accretion model.

As discussed above, the calculations reported here cover a wide
range of X-ray luminosities and not just the fiducial level in Table 1:
$\LX$  spans 3 orders of magnitude, from
$\LX=3\times 10^{29}$ to $\LX=3\times 10^{32}$ erg$\ps$.
This range may be compared with the X-ray luminosities in the Guedel et
al.~(2010) correlation study of the \NeII\ line, where the
measured X-ray luminosity varies over
the range $\sim 10^{29} - 10^{31}\, \ergps$. As we have just pointed out, it
is difficult to make a direct connection between our model parameter $\LX$
and measured X-ray luminosities. However, our main current goal is to deduce
the overall trend of line luminosities with X-ray luminosity and not detailed 
agreement for any particular object. This point is reinforced by the fact 
that the numerical results are affected by other model parameters which are not varied.
For example, the effect of X-ray ionization at the base of the wind is
affected by the location of our model point sources. Thus, if the elevated
source is moved up in $z$, a smaller $\LX$ would be required to generate the
same level of ionization.

\section{Results}

\begin{figure*}
\epsscale{1.0}
\plottwo{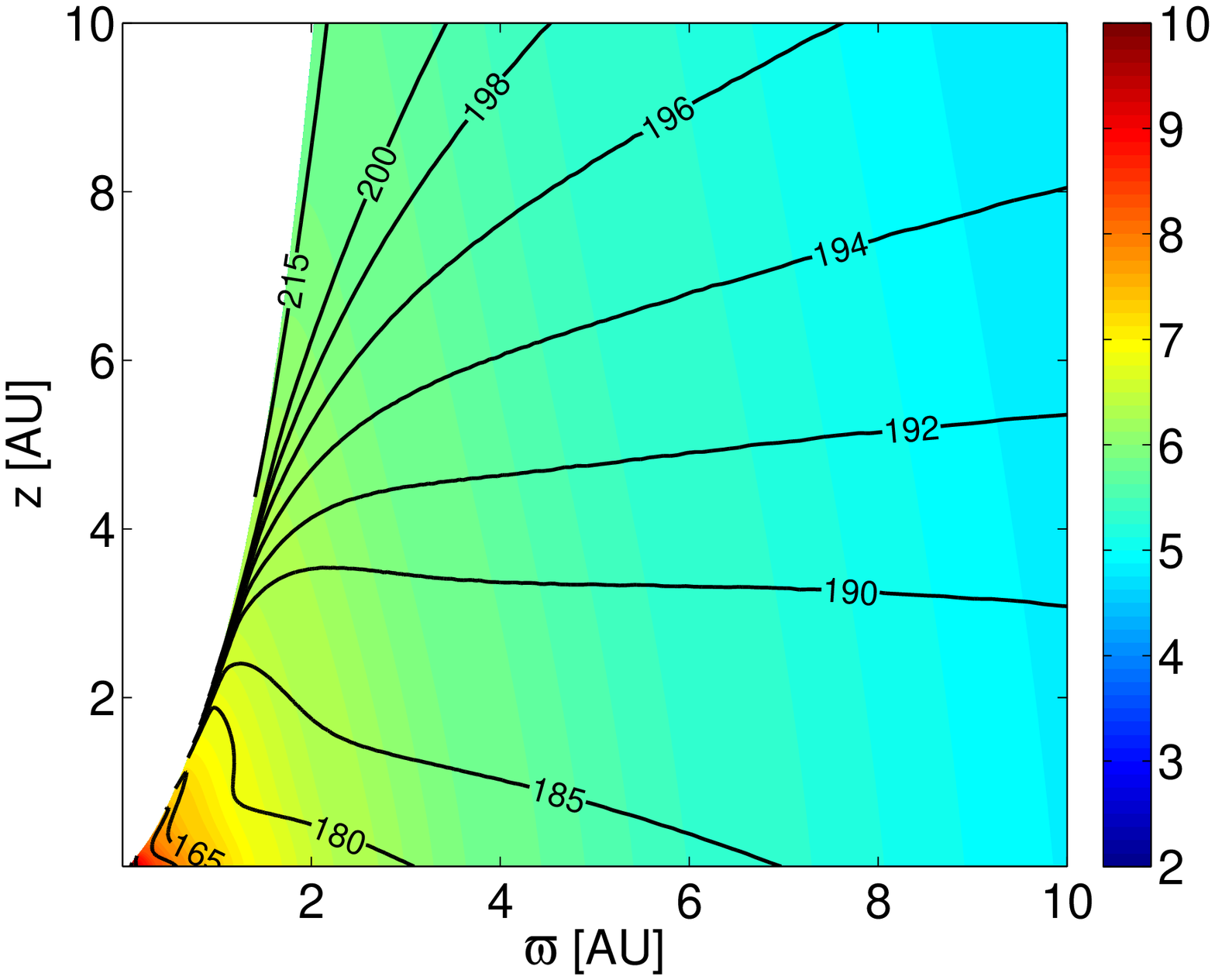}{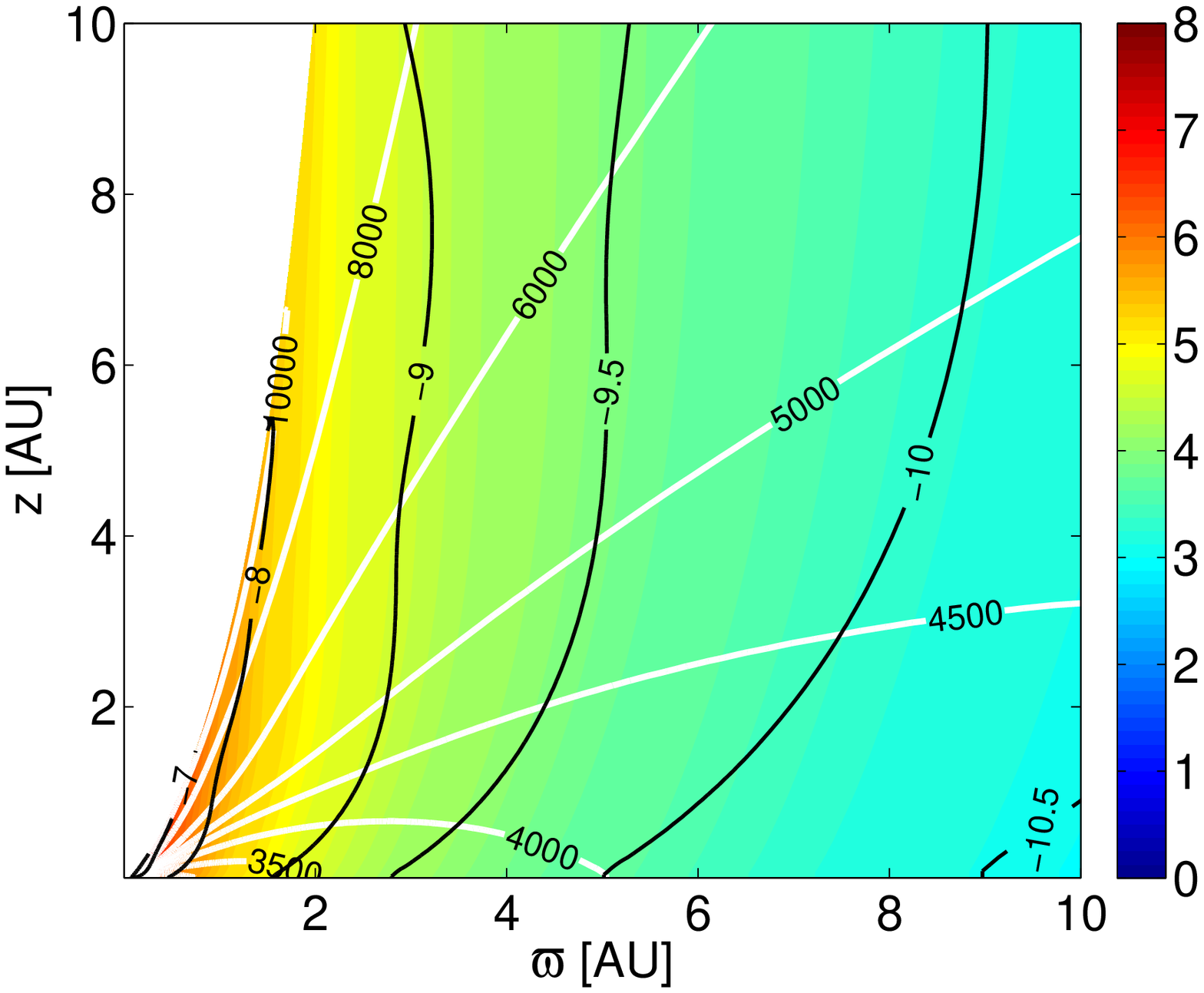}
\caption{(a) Background density of atomic hydrogen (color scale)
and velocity contours (solid black lines) for the X-wind reference case.
(b) Electron density (color scale), ionization rate $\zeta$
(solid black lines) and temperature (white lines). Volumetric densities
are given in cgs units and velocities in $\kmps$.}
\end{figure*}

Figure 1 shows the physical properties of the inner flow for the X-wind
reference model in Table~1.  The number density of atomic hydrogen is
displayed with a color scale in Figure~1(a); the ordinate is
the vertical distance $z$ and the abscissa is the cylindrical distance
$\varpi$ in units of AU. Beyond $\varpi = 5$\,AU, the density is
approximately collimated cylindrically. The velocity field (solid lines)
reaches near-terminal velocity within a few AU from launch and tends to
be radial at large distances. The occurrence of a cylindrically stratified
density in an expansive flow is characteristic of the magnetically
collimated jet in the X-wind model.

The electron density is important for the production of the neon lines
because the fine-structure transitions are excited by electron collisions.
Figure~1(b) shows the electron density in color and the X-ray
ionization rate $\zeta$ as solid black lines. Because the ionization
tends to be frozen in, the electron density in the inner jet decreases
more slowly with $\varpi$ than the ionization rate, which decreases rapidly
due to absorption and inverse-square dilution of the X-rays. Both the
electron density and the ionization rate manifest a conical shape
similar to that often seen near the base of optical jets. Isotherms are
also shown in Figure~1(b) as white lines. The magnitudes of the electron
density and the temperature are suitable for exciting the neon lines.
For example, the critical density of the \NeII\ 12.8\,$\micron$ line
is $4-5\times 10^5 \, \percc$ for temperatures in the range
$5,000-10,000$\,K. In the very inner part of the jet, $\varpi < 2$\,AU, the
upper level of the transition is populated close to thermal equilibrium,
but at larger distances the excitation is sub-thermal. For the reference
case, most of the neon is ionized with roughly equal fractions in Ne$^+$
and Ne$^{++}$; Ne$^{++}$ gradually gives way to neutral neon at large
distances along the innermost core of the jet, while it continues to
account for $\sim 50\%$ of neon in the wide-angle wind.

We have calculated the luminosity of the neon lines for a range of models 
in which the main parameters, the X-ray luminosity $\LX$ and the mass-loss 
rate $\Mw$ both for a one-sided jet are varied away from the 
reference values in Table 1. In some cases, their {\it ratio} is kept fixed at
${\LX/\Mw} = 3 \times 10^{13}\, {\rm erg}\,{\rm g}^{-1}$,
while the wind mass-loss rate changes from
$5\times 10^{-7}$ to $5\times 10^{-10}$ $M_\odot$\,yr$^{-1}$.
This range extends from a somewhat embedded Class I source down to the level
of a revealed classical TTS. Line fluxes are calculated for
the standard sequence of parameters used by Shang et al.~(2004). Then the
mass-loss rates are increased and decreased by factors of 3 keeping
${\LX}/{\Mw}$ fixed. Finally, the constraint of fixed
${\LX}/{\Mw}$ is removed for each of these standard cases, and $\LX$ 
is varied by 1 order of magnitude above and below the standard series.

The ionization balance of the neon species is calculated by the method given
in Section 2 of GNI07. Excitation by neutral hydrogen has been ignored because 
the large electron fractions make electron collisions dominant.
The total abundance of neon is set at $10^{-4}$ relative
to hydrogen and that of oxygen at $6.6\times 10^{-4}$, following the discussion
in GNI07.  The populations of the lowest five levels of Ne$^{++}$ and \OI\ were
solved exactly as were the two levels for Ne$^+$. The lines were assumed to
be optically thin, a good approximation for the region shown in Figure 1.

\begin{figure}
\epsscale{0.92}
\begin{center}\plotone{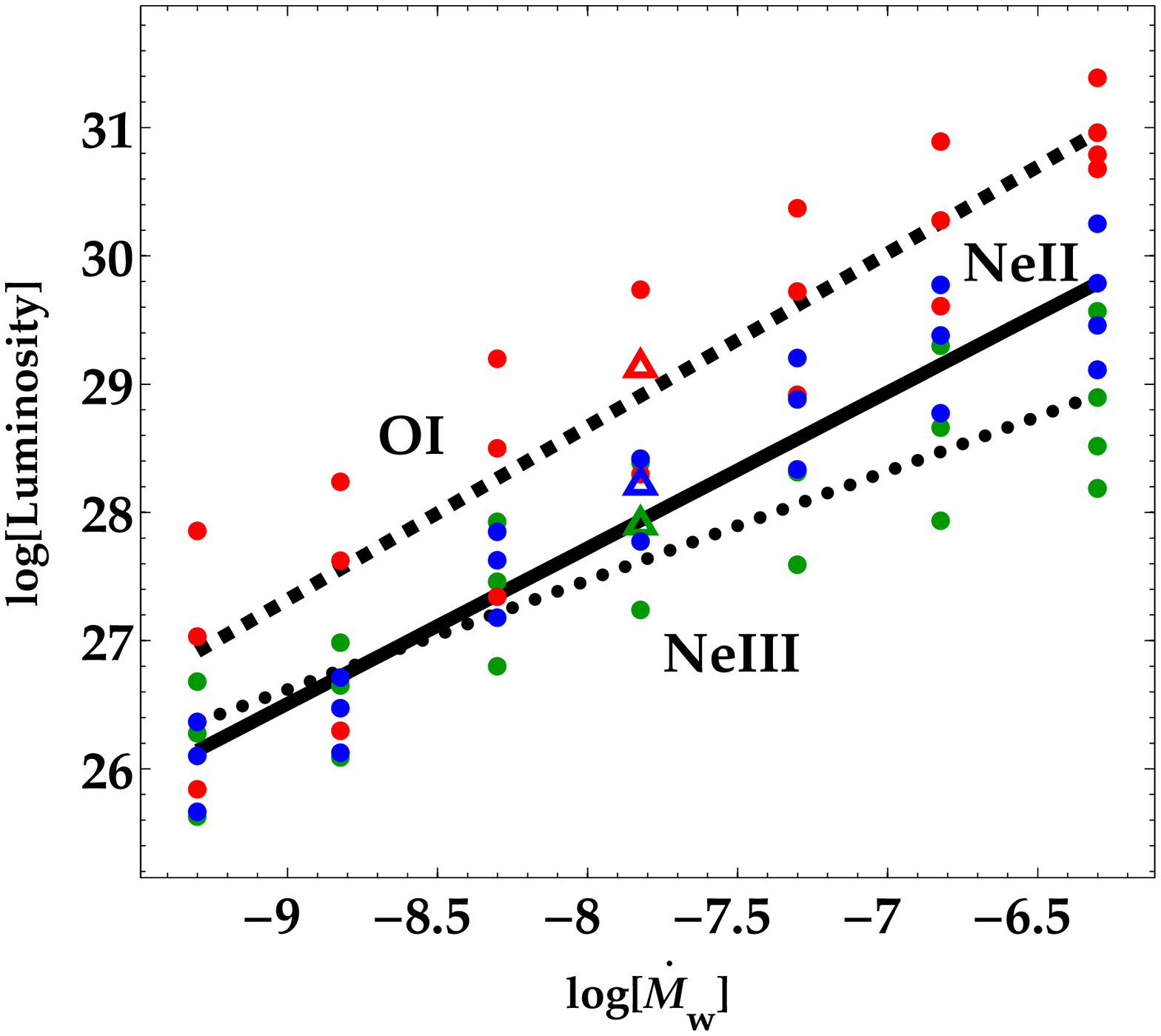}\\ \hspace{1cm}(a)\end{center}
\begin{center}\plotone{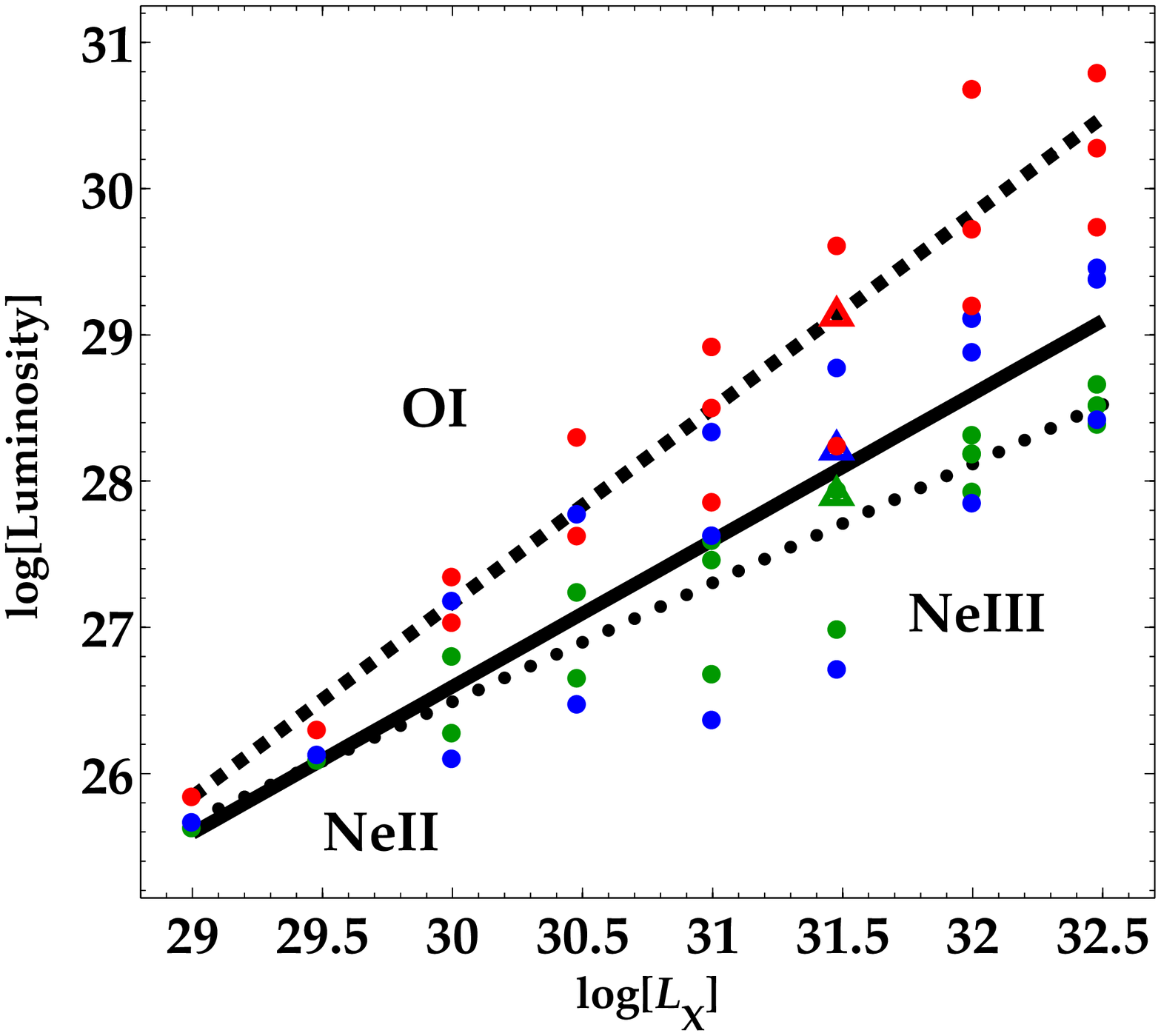}\\ \hspace{1cm}(b)\end{center}
\begin{center}\plotone{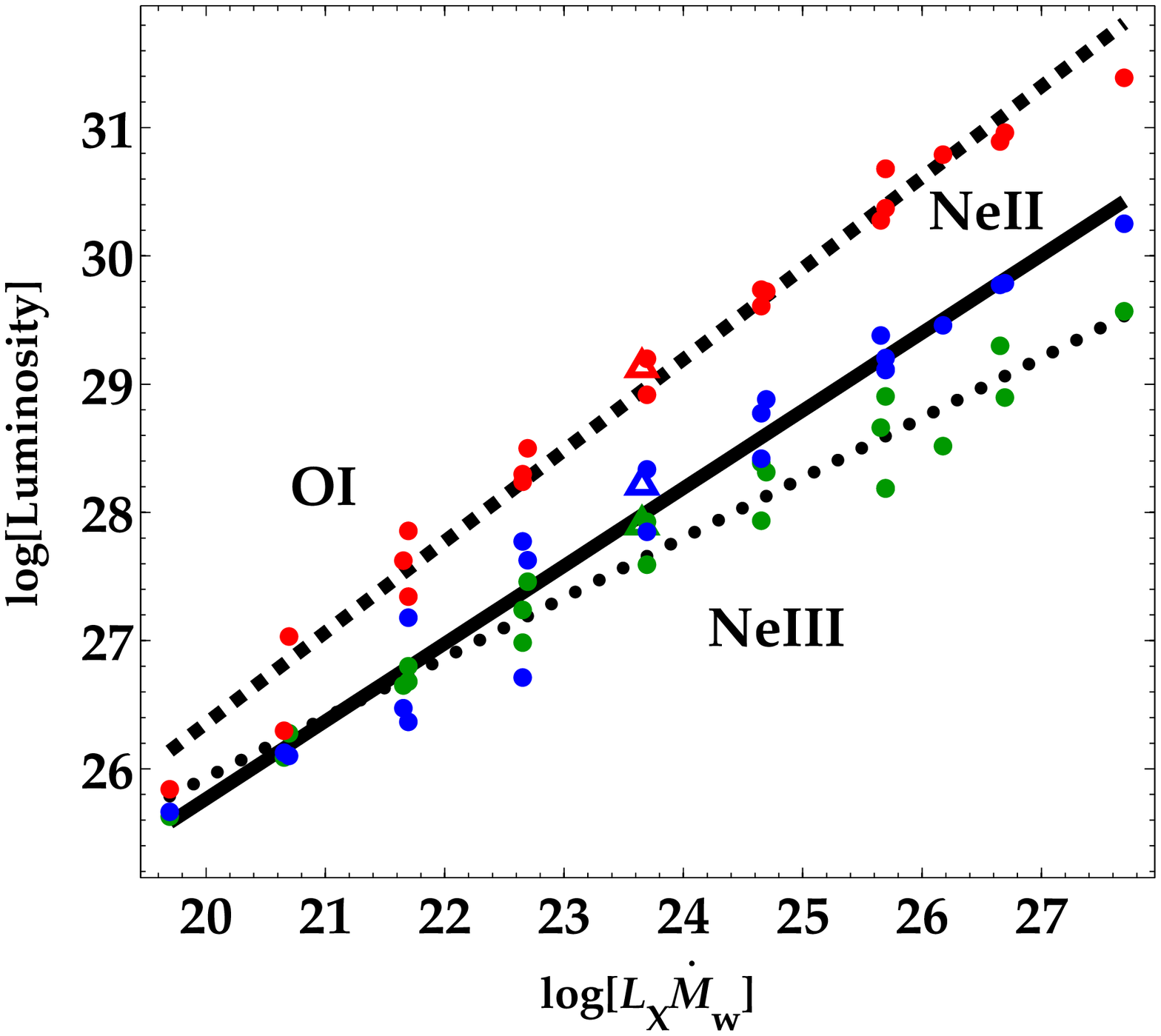}\\ \hspace{1cm}(c)\end{center}
\caption{(a) Line luminosities
\NeII\ 12.8 $\micron$ (solid line),
\NeIII\ 15.55 $\micron$ (dotted line), and the
\OI\ $\lambda$6300 (dashed line) from a one-sided jet, plotted versus
$\Mw$ in units of $M_\odot$\,yr$^{-1}$.
The vertical axis labels line luminosities in erg s$^{-1}$.
The open triangles mark the reference case of Table 1.
$\Mw$ is the mass-loss rate from one hemisphere.
The slopes of the power-law fits are: $1.351 \pm 0.132$ (\OI);
$1.215 \pm 0.084$ (\NeII); $0.851 \pm 0.110$ (\NeIII).
(b) Same as (a), except the line luminosities are plotted versus
the X-Ray luminosity $\LX$ of a one-sided jet in units of erg s$^{-1}$.
The slopes of the power-law fits are: $1.328 \pm 0.117$ (\OI);
$1.001\pm 0.152$ (\NeII); $0.813 \pm 0.068$ (\NeIII).
(c) Same as (a), except the line luminosities are plotted versus
$\LX\Mw$ in units of erg s$^{-1}$ $\times$ $M_\odot$\,yr$^{-1}$.
The slopes of the power-law fits are: $0.709 \pm 0.022$ (\OI);
$0.606 \pm 0.026$ (\NeII); $0.468 \pm 0.022$ (\NeIII).}
\end{figure}

Figure 2 plots the luminosity of the \NeII\ 12.8\,$\mu$m and \NeIII\
15.55\,$\mu$m lines and the forbidden transitions of \OI\ $\lambda6300$
against three variables: (a) the mass-loss rate $\Mw$, (b) the X-ray 
luminosity $\LX$, and (c) the product $\LX\Mw$ --- introduced by 
Guedel et al.~(2008) in a search for empirical correlations of the \NeII\ 
luminosity with other properties of YSOs. Values from both groups of 
calculations (fixed and variable $\LX/\Mw$) are plotted. The fluctuations 
in Figures~2(a) and (b) show the effects of the variable not plotted. 
The quantity $\Mw$ is intrinsic to X-wind theory, whereas $\LX$ is an external 
parameter that was added in the development of the physical properties of the 
X-wind by SGSL. There is no self-consistent theory available yet that treats both
the magnetospheric wind and accretion generation together with X-ray
emission. The heating coefficient $\alpha_h$ in Equation~\ref{mech_heat} is
another external parameter introduced by SGSL, but it is less effective in
changing the physical properties of the outflow, and for purposes of
simplicity it is not varied very far from the value in Table~1 for the
reference model. The large open triangles in Figure 2 are for the reference
case. The model calculations all fit power laws with relatively small scatter,
i.e., the $3\sigma$ uncertainties in the least-square slopes are all
significantly smaller than the slopes themselves. The parameter $\alpha_h$ has
been kept fixed at 0.002 in Figure~2; varying $\alpha_h$ by a factor of a few
would naturally introduce some scatter without changing the slope. Unlike the
case of disk emission (MGN08), the correlation with X-rays in Figure~2(b) is
reasonably well founded because the effects of the other main parameter
$\Mw$ are seen in Figure~2(b) to be relatively modest.\footnote{Figure 2(b)
is similar to Figure~16 of MGN08, where the luminosity of several diagnostic
lines, including the \NeII\ 12.8\,$\mu$m line, are plotted versus $\LX$ for
the same disk structure, that of a generic TTS disk. A correlation diagram
for a real sample of observed TTS disks may not follow this power law because
it will likely include a variety of disks at different ages and density
structures.}

Somewhat surprisingly, the fits with the empirical parameter
$\LX\Mw$ in Figure~2 are superior to the other two. The ratio of the
$1\sigma$ uncertainty in slope to the slope itself is $\sim 0.05$. A
possible interpretation of this correlation is that it reflects the
primary role of the electron density in producing the neon ions and in
exciting their line emission. This in turn follows from the idea that
$\LX/\Mw$ is the global average of the local X-ray radiation
parameter, $\zeta/n$. This quantity is roughly proportional to the square
of the electron fraction, so that the square root of the empirical
parameter, $\LX\Mw$, is a measure of the {\it electron density.}

The slope of the correlations for \OI\ in Figure~2 is larger than those of
the \NeII\ and \NeIII\ lines, and the \NeII\ slope may be slightly larger
than the one for \NeIII. These differences can be traced to the ionization
fractions of these ions. As pointed out by MGN08, direct ionization of neutral
oxygen by X-rays is not competitive compared with charge exchange with H$^+$.
As a result, the neutral oxygen fraction in the present calculations
is above 90\% except for very small values of $\LX$ and $\Mw$. On the
other hand, the Ne$^+$ and Ne$^{++}$ fractions are sensitive to the local 
X-ray ionization parameter, $\zeta/n$. Thus for the reference case, the
Ne$^{++}$/Ne$^+$ ratio approaches unity in the inner jet, but then falls to
$\sim 0.5$ throughout the rest of inner region of the outflow
($\varpi < 10$\,AU) due to attenuation and inverse-square dilution of the 
X-ray flux. For larger values of $\Mw$, attenuation takes an increasing toll
and reduces the Ne$^{++}$/Ne$^+$ ratio, although it is always larger in the
inner jet than in the outflow at large. This explains the tendency of the 
luminosity ratio $L($\NeIII$)/L($\NeII$)$ in Figures~2(a) and (c) to
decrease with increasing $\Mw$. This tendency is at most moderately significant 
because of the large fluctuations in the theoretical predictions
of the luminosities of the \NeII\ and \NeIII\ lines in Figure~2.

\begin{figure}
\epsscale{1.0}
\plotone{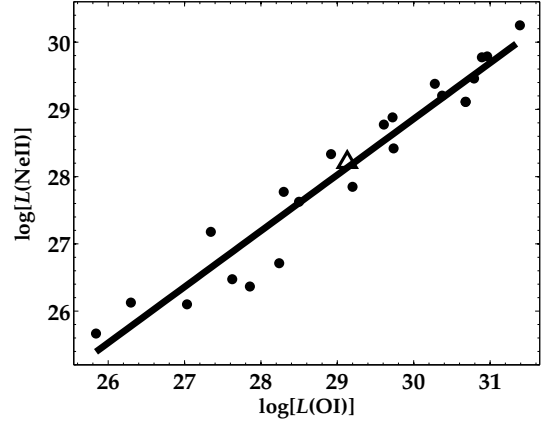}
\caption{Luminosity of the \NeII\ 12.8 $\micron$ line vs.~the \OI\ $\lambda$6300
luminosity in a log-log plot; the units are erg s$^{-1}$.
The correlation line has a slope close to 0.8.}
\end{figure}

The fact that the luminosities in Figure~2 correlate well with the variables
$\Mw$ and $\LX$ means that they correlate with one another. This is illustrated 
in Figure 3, where the \NeII\ 12.8 $\micron$ luminosity is plotted against the
\OI\ $\lambda$6300 luminosity on a log-log scale. The slope of the correlation
line is $0.83 \pm 0.048$. The \OI\ $\lambda$6300 lines have long been used as 
diagnostics of YSO jets. The predicted correlation with \NeII\ suggests that
the latter line may also be a good diagnostic for these jets. Because our
calculations are based on X-wind theory supplemented by chemical-physical 
considerations, including X-ray ionization of the outflow, Figure~3 offers
a potentially useful observational test of the excitation theory in SGSL
without the uncertainties of external parameters such as $\LX$.
A preliminary inspection of the available data for jet sources
in Guedel et al.~(2010) suggests a good correlation with our theory.
Further detailed comparisons of the \NeII\ 12.8 $\micron$ and the \OI\ $\lambda$6300 
luminosities should be helpful in understanding the origin of these
emission lines.

Figure 4(a) shows normalized profiles of the \NeII\ 12.8 $\micron$,
\NeIII\ 15.55 $\micron$ and the \OI\ $\lambda$6300 lines for the reference
case of Table 1 at an inclination angle of $45 \arcdeg$.
The calculations are for one side of the outflow and the results
have been binned to a spectral resolution of 1 $\kmps$. The predicted
emission extends over a broad range of velocities from $\sim -200$ to
$\sim +100\, \kmps$ at this inclination, The dominant feature is the large
blue-shifted peak close to the projected terminal velocity of the jet.
Figure 4(b) shows the contributions to the \NeII\ line shape as a function
of the cylindrical coordinate $\varpi$. As $\varpi$ increases, the height
of the peak grows from contributions of regions with ever-increasing radius
and outflow velocities close to the terminal value. The \NeII\ line
has a broader low-velocity wing than the other lines, indicating that \NeII\
fraction is more sensitive to the physical conditions close to the source and
the acceleration zone. On a larger scale, the \NeII\ line is generated at
relatively small distances, $\varpi < 5$\,AU, with the broad wing from
$\sim -50$ to $\sim +50\, \kmps$ produced within $\sim 1-2$\,AU of the axis.
This is close to the source of the wind; Figure~1(a) shows that the wind
velocity is already quite large by $z = 0.5$\,AU.

The line profiles vary with the inclination angle, which is important
for making comparisons with observations. Figures 5(a) and (b) show line shapes
at $30\arcdeg$ and $60\arcdeg$ for one side of the jet; they are similar to
the $45\arcdeg$ case shown in Figure 4(a). Significant changes occur
for inclinations near $90\arcdeg$ and $0\arcdeg$. Near these extremes, the
emission peak occurs close to the real terminal velocity for the face-on
case ($0\arcdeg$), while it approaches the stellar velocity for the edge-on
case ($90\arcdeg$). The X-wind jet flow has the form of a cylindrically symmetric
rounded cone or bowl. The peak velocity in the emission is determined by the 
projection of the terminal velocity. The line shape varies gradually with
inclination angle as the projected velocity peak moves toward larger or
smaller projected values. The low velocity wing remains broad and extends
to negative velocities. In theory, the jet is bipolar, and the full profiles
are in principle symmetric with respect to the stellar velocity. The receding
jet produces a red-shifted emission peak at the same magnitude of projected
velocity as the blue-shifted peak, and its low-velocity wing extends
to the blue. Depending on the actual viewing angle and orientation of the
system, the line profile of a jet with two visible lobes will be a mixture
of shapes from each side. If for some reason only the receding side is
visible, the profile will be peaked at the high red-shifted velocity.
Several authors have suggested that sources with observed high \NeII\ luminosity
are associated with jets (e.g., Guedel et al.~2010; van Boekel et al.~2009; 
Flaccomio et al.~2009). Our results support this view, and our specific
predictions can be checked by the detection of strong blue-shifted
line emission, as seen in Figures~4 and 5.

\begin{figure*}
\epsscale{0.8}
\plottwo{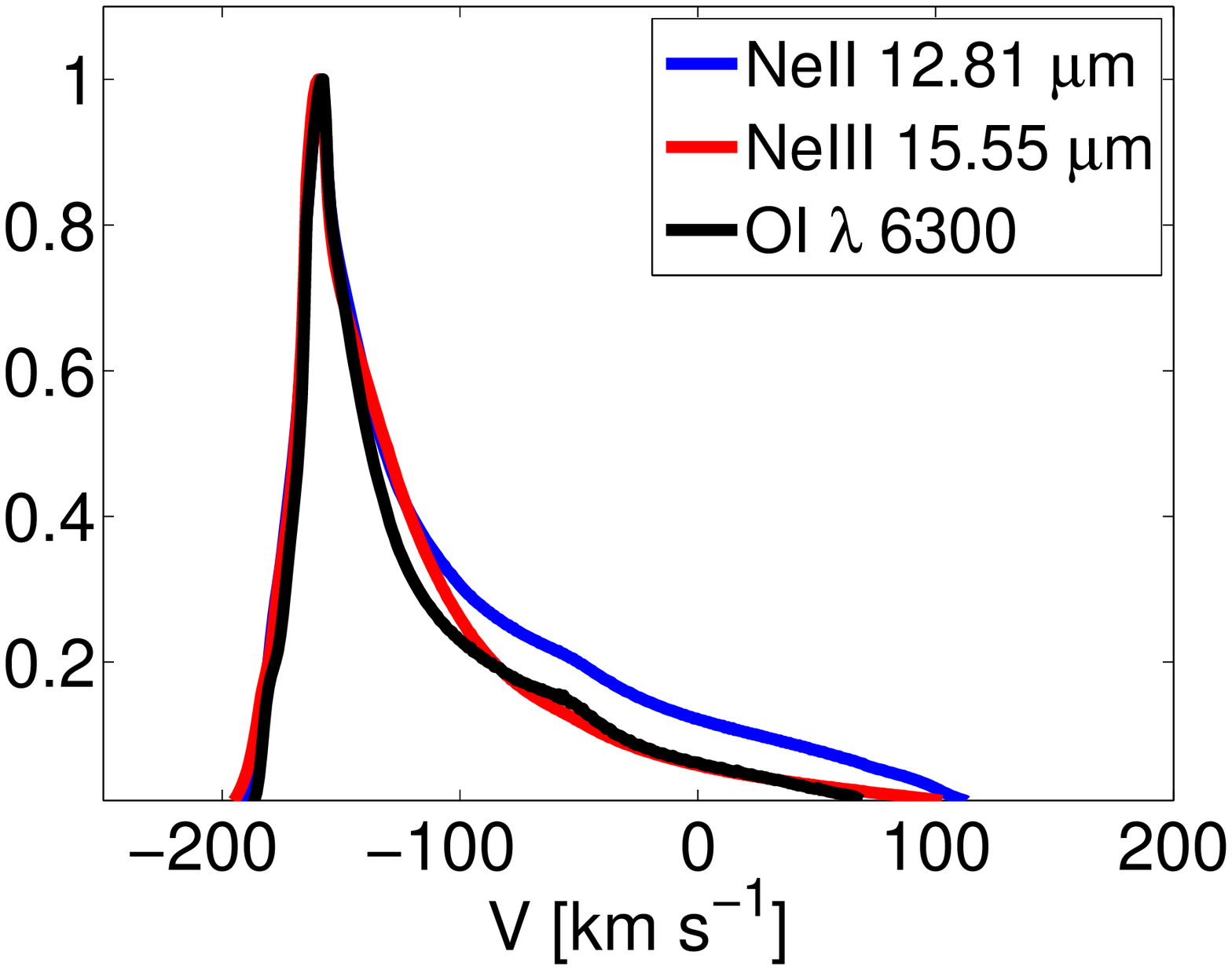}{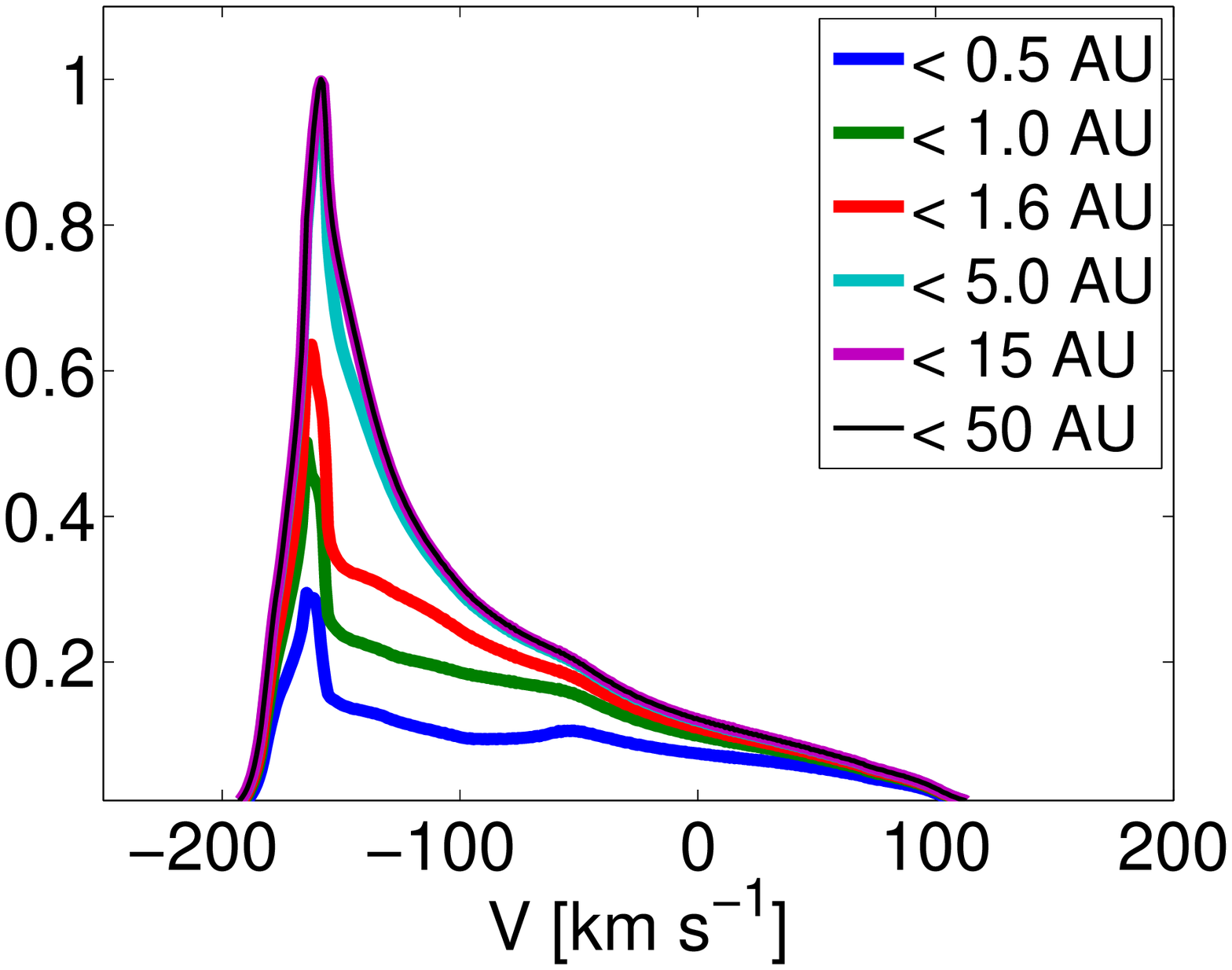}\\
(a)\hspace{7.8cm}(b)
\caption{(a) Line profiles for \NeII\ 12.8 $\micron$ (blue),
\NeIII\ 15.55 $\micron$ line (red) and the \OI\ $\lambda$6300 (black)
for the X-wind jet, calculated for the reference case of Table 1.
The inclination angle is $ 45 \arcdeg $, and the line shapes are normalized
to their peak values.
(b) Buildup of the \NeII\ line shape as a function of cylindrical radius $\varpi$ 
for the case of $45 \arcdeg$ inclination angle.}
\end{figure*}

\begin{figure}
\epsscale{0.8}
\begin{center}\plotone{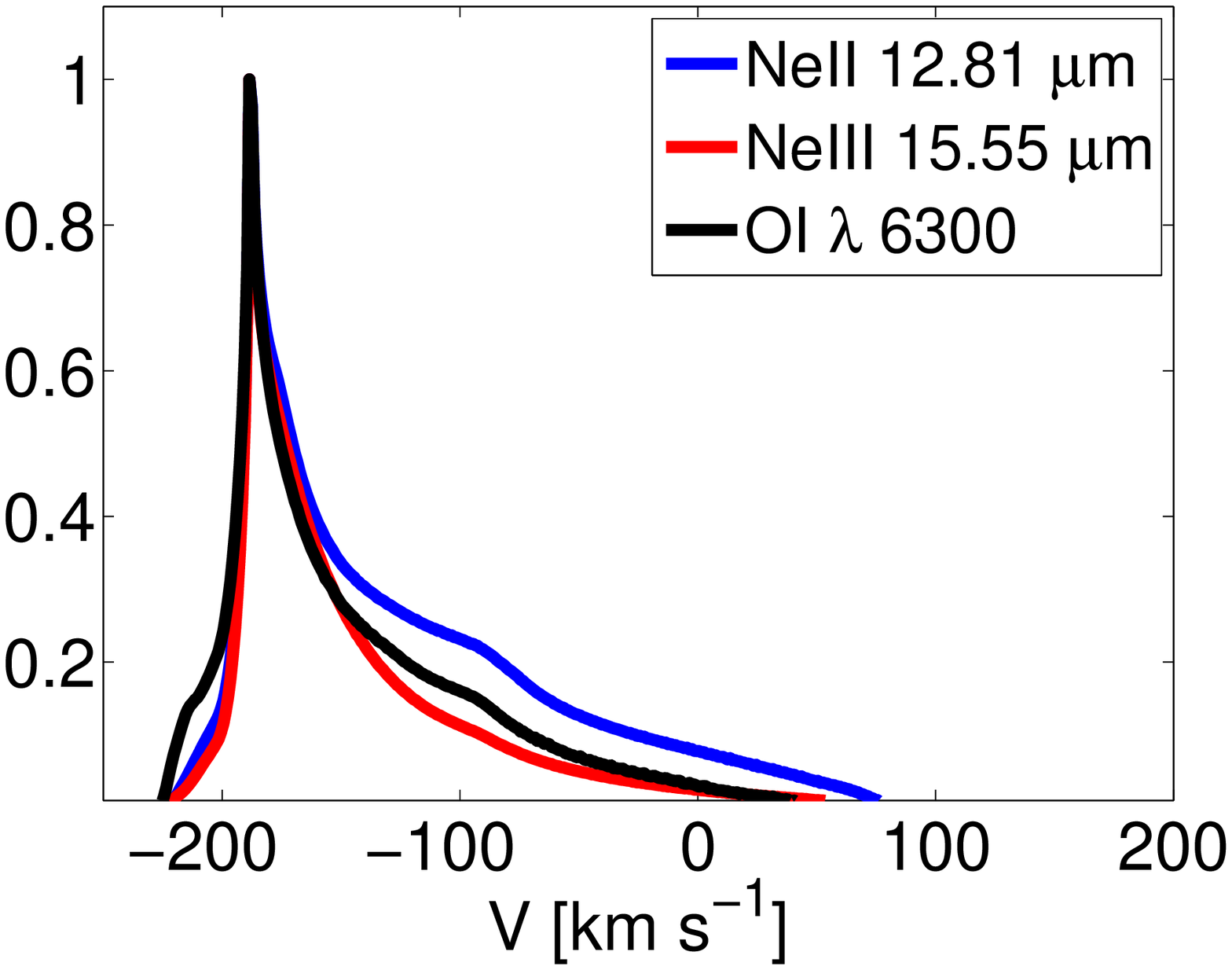}\\ \hspace{0.6cm}(a)\end{center}
\begin{center}\plotone{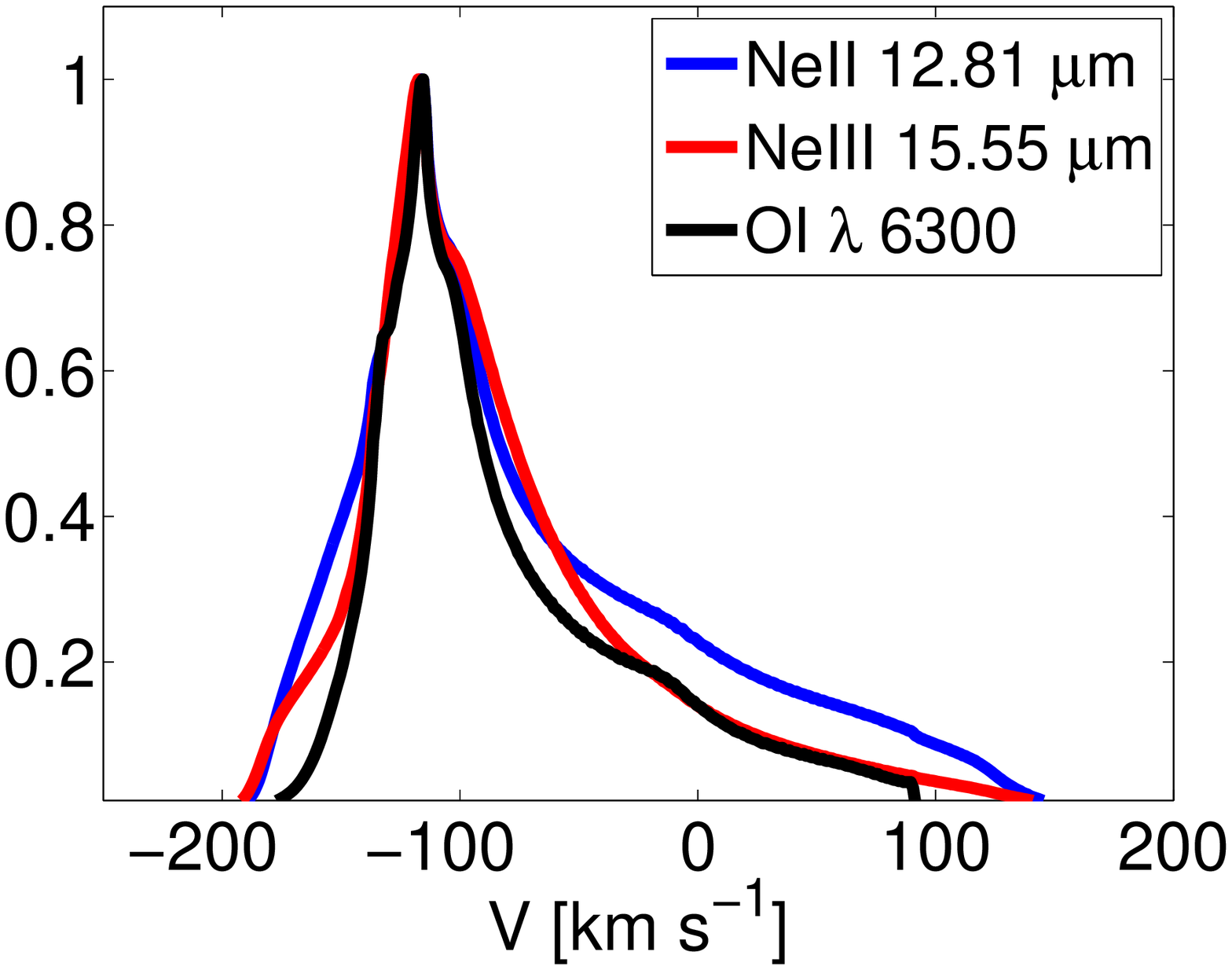}\\ \hspace{0.6cm}(b)\end{center}
\caption{(a) Line profiles for \NeII\ 12.8 $\micron$ (blue),
\NeIII\ 15.55 $\micron$ line (red) and the \OI\ $\lambda$6300 (black)
for the approaching X-wind jet, calculated for the reference case of Table 1.
The inclination angle is $ 30 \arcdeg $, and the line shapes are normalized
to their peak values.
(b) Line profiles for \NeII\ 12.8 $\micron$ (blue),
\NeIII\ 15.55 $\micron$ line (red) and the \OI\ $\lambda$6300 (black)
for the X-wind jet, calculated for the reference case of Table 1.
The inclination angle is $ 60 \arcdeg $, and the line shapes are normalized
to their peak values.}
\end{figure}

\section{Discussion and Conclusions}

We have calculated the emission of the \NeII\ 12.8 $\micron$, \NeIII\ 15.55
$\micron$ and \OI\ $\lambda$6300 lines for the X-wind model for the outflows
from YSOs. The oxygen lines are well-established diagnostics of jets, whereas
the \NeII\ 12.8 $\micron$ line has only recently become a probe of the
circumstellar gas of young stars, as described in the Introduction. We have
calculated the dependence of the line emission on the two main parameters of
the model, the wind mass-loss rate $\Mw$ and the X-ray luminosity $\LX$.
The mass-loss rate is an intrinsic parameter of X-wind theory, whereas the
X-ray luminosity was added by SGSL in modeling the physical properties of
the wind. Figure 2(a) expresses our first basic result. It shows that the
line luminosities correlate well with the X-wind parameter $\Mw$. Figure~2(a)
also shows that, for mass-loss rates in excess of those characteristic
of Class II TTSs, the luminosity of the \NeII\ 12.8 $\micron$ is well above
the observed range for this line given by Guedel et al.~(2010). It
is interesting that the combination parameter, $\LX\Mw$, gives the best 
correlation of the three shown in Figure~2. Figure 3 gives another example of 
a good correlation, in this case between the \OI\ forbidden transitions and 
the \NeII\ fine-structure line. It suggests that the \NeII\ 12.8 $\micron$ 
line is an excellent probe of the jets from YSOs,
as are of course the well-established \OI\ $\lambda$6300 line.

The theoretical line profiles displayed in Figures~4 and 5 have a distinctive shape.
For an approaching one-sided jet, they have a strong blue-shifted
peak near the wind terminal velocity that is associated mainly with the inner
part of the jet within $\varpi < 5$\,AU. There is also a broad shoulder centered
around the stellar velocity that arises near the source of the jet and the
acceleration region of the outflow. These features reflect some of the unique
properties of the X-wind model of outflows from YSOs. Observations to test these 
predictions for high luminosity sources are feasible for both the \NeII\ 12.8
$\mu$m and the \OI\ $\lambda6300$ lines. High-spectral resolution measurements
would be of great interest for understanding the source of the \NeII\ and \OI\
line emission, the role of X-rays in determining the physical properties of
outflows, and the dynamics of jet formation. The blue-shifted peak of the
\NeII\ line is probably the most important signature of jet emission, because
the low-velocity shoulder may be contaminated in some cases by emission from
the X-ray irradiated disk, either from the disk proper (MGN08) or from a 
photo-evaporated wind (Owen et al.~2010; Ercolano \& Owen 2010).

Line profile measurements have already been carried out for seven low-mass YSOs, 
as discussed in the Introduction. Four of the seven are transition objects which 
have evolved disks with substantial gaps in the inner dust distribution (TW Hya, 
GM Aur, CS Cha, and T Cha), typically on the AU scale. These four objects all show 
clear signatures of accretion onto the stellar surface, which implies that there 
is a significant amount of gas present within the inner radius of the dust gap. 
These are not the most appropriate systems for applying the X-wind model, 
originally conceived to describe YSOs with strong outflows. Two of the seven YSOs 
(T Tau and Sz 73) are outflow sources. The simple X-wind model used in this paper 
is inadequate to deal with the T Tau triple stellar system, which has multiple 
sources of X-rays and \NeII\ line emission. Of all the line shapes reported by 
Pascucci \& Sterzik (2009), the broad, blue-shifted line of Sz 73 is closest to 
the profiles we show in Figures~4 and 5. However, little is known about this YSO, 
and it would be premature to conclude that it agrees (or disagrees) with our model 
until more information is obtained, e.g, its X-ray properties determined, and 
detailed modeling is carried out. This leaves AA Tau, whose line shape has been 
measured by Najita et al.~(2009) to be broad (FWHM $\sim 70\, \kmps$) and with a 
red-shifted peak at $+15\, \kmps$. There is also emission corresponding to the 
blue-shifted velocity, but the spectrum is noisy and difficult to characterize. 
With its large inclination angle, it is possible that higher signal-to-noise 
observations may show the line shape expected on the basis of our X-wind calculations, 
especially since AA Tau manifests some evidence for outflow (Bouvier et al.~2003, 2007). 
On the other hand, the flux of its \NeII\ 12.8 $\micron$ line is at the level obtained 
for X-ray irradiated disk models (MGN08). Thus any serious model calculation would 
have to include all the sources of emission, the accretion funnel, the disk and the wind. 
We would like to recommend that searches be made for \NeII\ 12.8 $\micron$ line emission 
from strong jets already detected in forbidden transitions from \OI\ and other heavy 
ions using high-resolution MIR spectrometers on large telescopes.

Hollenbach \& Gorti (2009) have provided an alternative view of the
\NeII\ 12.8 $\micron$ line emission for sources with high accretion rates.
In their theory, the main source of ionization of neon and the excitation
of the 12.8 $\micron$ line is high-velocity ($\sim 100\, \kmps$) shocks
that affect the entire jet. While high-velocity shocks may generate large
enough temperatures to ionize neon in localized regions, there is little
observational evidence for the strong and continuous shocking of jets. The bulk of the
magnetized flow is at most mildly shocked, despite the high terminal
speeds reached close to the source. It is these mild shocks that are
responsible for heating the outflow in the present model. Hollenbach \&
Gorti (2009) do not calculate line shapes for their high-velocity shock
model. Although they are likely to be broad ($\sim 100\, \kmps$), the line
shapes can be expected to be different from the ones shown here because
the broadening arises downstream along the jet where the strong shocks
occur. In contrast, most of the \NeII\ and \NeIII\ emission in our work is
generated within or close to the acceleration region that surrounds
the source, so that the line shape takes its final form close to the
source, as shown in Figure~4.

Guedel et al.~(2010) have made extensive searches for observational
correlations of the \NeII\ 12.8 $\micron$ line luminosity with X-ray
luminosity and other properties of YSOs, as have Flaccomio et al.~(2009).
They do not find any particular trend for low-accretion systems, where
the measured \NeII\ luminosities span a range of a factor of 10 or more
about a median of approximately $L($\NeII$) \approxeq 5 \times
10^{-6}\, L_{\sun}$ (or $ \log L($\NeII$) = 28.25$ in cgs units).
This value is close to our reference theoretical value, the large open
triangle in Figure~2. It is also similar to the value predicted by MGN08
for an X-ray irradiated disk of a typical TTS, although the \NeII\ emission
from disks has not been calculated for any other case. Calculations
for photo-evaporated winds give similar values (Alexander 2008; Ercolano
\& Owen 2010). One explanation for the large number of measurements for 
both optically thick and transition disks with this level of \NeII\ 
luminosity is that other variables are important, and not just the X-ray 
luminosity. Schisano et al.~(2010) illustrate this possibility by showing 
that variations in disk flaring and X-ray spectrum give rise to a scatter 
in the predicted \NeII\ luminosity similar to what is observed. According 
to our calculations, outflows may also contribute in some cases, but they 
are not the dominant source of emission for low or even moderate accretion 
rates and X-ray luminosities. However, as shown in Figure~2, they may 
dominate for high mass-loss rates and X-ray luminosities.
Thus our calculations support the ``bi-modal'' picture of Guedel et al.~(2010)
and van Boekel et al.~(2009), especially their suggestion that the \NeII\
luminosity of high-accretion sources arises in outflows.

On the basis of a regression analysis, Guedel et al.~(2010) also concluded
that the \NeII\ luminosities of jet sources correlate, not only with X-ray
luminosity, but with accretion rate and mass-loss rate. The significance of
these correlations is diminished by the relatively small sample of sources
for which the parameters have been measured and by scatter in the data.
This is especially true in the case of the mass-loss rate, where only two
sources (T Tau N and DG Tau) have exceptionally large \NeII\ luminosities.
Without relying on these two ``outliers'',\footnote{T Tau N is an outlier
in that it is part of a compact triple system; DG Tau has an unusually small
{\it observed} X-ray luminosity.} the correlation of the
\NeII\ luminosity with mass-loss rate is not yet definitively established.
Such a correlation would be directly relevant to the present calculations
because, in X-wind theory, it is the mass-loss rate that directly determines
the properties of the outflow, rather than the accretion rate, although the
two are related. Mass-loss rates of YSO jets are often determined from bright
optical emission lines such as \OI\ $\lambda$6300 (e.g., Hartigan et al.~1994,
1995), but the results for YSOs of the same mass can range over several
dex (White \& Hillenbrand 2004). In light of such uncertainties, it would
be better to compare the luminosity of the \OI\ $\lambda$6300 with that of
the \NeII\ line, as we do in Figure~3. Guedel et al.~(2010) test this
correlation with data for sources they identify as optically thick disks
without jets, transition disks, and jet sources. A preliminary inspection
of the available data for the seven jet sources in their Figure~4 with
$L($\NeII$) > 8 \times 10^{-6}\, L_{\sun}$ suggests a good correlation
with our theory. Further observations of the \NeII\ and \OI\ emission from
strong jets including line shape measurements would be extremely useful in
this connection.

The correlation predictions in Figure 2 suggest another possible test of
our calculations using the \NeIII\ 15.55\,$\mu$m line, although only a few
detections of this line have been reported so far. In Figure~2(a) the
predicted but uncertain ratio of the \NeII\ 12.8\,$\mu$m to the
\NeIII\ 15.55\,$\mu$m line flux is in the range $\sim 1-10$; the \NeIII\ flux
is $\sim 10^{-14}-10^{-13}\, {\rm erg}\,{\rm cm}^{-2}\,{\rm s}^{-1}$. Lahuis et
al.~(2007) reported a tentative detection of the \NeIII\ line with {\it Spitzer}
for Sz 102, a poorly studied source with a prominent jet, known to emit
primarily soft X-rays (Guedel et al.~2010). Lahuis et al.~(2007) give the neon line
fluxes as
$L($\NeII$) = 3.6 \times 10^{-14}\, {\rm erg}\,{\rm cm}^{-2}\,{\rm s}^{-1}$
and
$L($\NeIII$) = 2.3 \times 10^{-15}\, {\rm erg}\,{\rm cm}^{-2}\,{\rm s}^{-1}$,
or a \NeIII/\NeII\ luminosity ratio of $\sim 1/16$. This small value is consistent
with a disk origin for the neon lines, but the measured fluxes are smaller
than predicted by MGN08 for the generic TTS disk and the measured X-ray
luminosity. Flaccomio et al.~(2009) detected both neon lines from the F7 Class III
YSO WL5/GY246 in the $\rho$ Oph cluster. Although the luminosities are
in rough accord with a disk origin, the \NeIII/\NeII\ luminosity ratio of
$\sim 1/4$ is somewhat higher than given by MGN08 and more like the values
obtained here for jets. However, we would not expect to find a strong jet
in a Class III YSO, so the disk model is favored in this case. A careful
determination of upper limits if not detections in {\it Spitzer} spectra of
high luminosity sources of the \NeIII\ 15.55\,$\mu$m line could provide
further checks of our theory. Since {\it Spitzer} operations have entered
the warm phase, further direct detections of this line from space will not
be available for some time. However the optical lines of \NeIII\ $\lambda$3869/3967
are accessible from the ground, and they have been observed from H\,{\sc ii} regions
and other highly-ionized sources. Preliminary calculations indicate that
these lines have about the same luminosity as the \NeIII\ 15.55\,$\mu$m line.
We suggest that these lines be used to further probe the ionization state and
emission characteristics of jets and disks around low-mass YSOs.

In conclusion, we have calculated the emission of the fine-structure lines
of neon and of the forbidden \OI\ transitions near 6300\,\AA, according to
the X-wind model of jets. These lines trace similar aspects of the outflow.
They correlate well with the X-wind model parameters and with the
combination variable $\LX\Mw$, as well as with one another. We have
suggested several tests of the calculations, including the measurement of the
unique blue-shifted peak in the line profile near the wind terminal velocity.
We also suggest that jets may account for the luminous mode in the
bimodal picture of \NeII\ correlations proposed by Guedel et al.~(2010) and
van Boekel et al.~(2009).

\acknowledgments

The authors thank Barbara Ercolano and Joan Najita for helpful discussions and
careful reading of the manuscript and the referee for helpful comments and suggestions.
They acknowledge support from the National Science Council of Taiwan through
grants NSC97-2112-M-001-018-MY3 and NSC96-2752-M-001-001-PAE to the Theoretical
Institute for Advanced Research in Astrophysics (TIARA) under the Excellence of
Research Program. This work has been supported at Berkeley by the NSF grant
AST-0507423 and the NASA grant NNG06GF88G.

\end{document}